\documentclass[a4paper,11pt]{article}
\usepackage{amsmath, amssymb}
\usepackage{color, bm}
\usepackage{cite}

\newcommand{\email}[1]{\footnote{email:#1}}

\DeclareMathOperator{\rk}{rk}
\DeclareMathOperator{\Tr}{Tr}
\DeclareMathOperator{\tr}{tr}

\newcommand{\g}{{\sf g}}
\newcommand{\h}{{\sf h}}
\newcommand{\K}{{\rm K3}}

\renewcommand{\O}{{\cal O}}
\renewcommand{\P}{{\mathbb P}}
\newcommand{\Z}{{\mathbb Z}}

\newcommand{\be}{\begin{equation}}
\newcommand{\ee}{\end{equation}}

\newcommand{\red}{\color{black}}

\begin{document}

\title{Small-Instanton Transitions in F-theory}
\author{Stephen Angus$^{\rm a}$\email{sangus@ewha.ac.kr} \ and Kang-Sin Choi$^{\rm b}$\email{kangsin@ewha.ac.kr}
 \\ \it \normalsize $^{\rm a}$ Institute of Mathematical Sciences, Ewha Womans University, Seoul 03760, Korea \\
\it \normalsize $^{\rm b}$ Scranton Honors Program, Ewha Womans University, Seoul 03760, Korea}
\date{}
\maketitle

\begin{abstract}
\noindent
We study the phase transition between $G$-instantons and D3-branes. A $G$-instanton is a classical solution to the self-dual equation of the M/F-theory four-form field strength $G$ in the complex fourfold. This phase transition is dual to that between small instantons and 5-branes in the heterotic string. Using $G$ as a background gauge flux, we may dynamically control the gauge symmetry breaking, connect between different vacua of F-theory and understand D3-branes in terms of group-theoretical quantities. We also discuss the resulting chirality change and preservation of anomaly freedom.
\end{abstract}

\newpage

\section{Introduction and summary}

String theory provides a framework for completion of the Standard Model. 
The global consistency conditions of string theory, principally arising from the one-loop vacuum-to-vacuum amplitude, are more fundamental than those of field theory.  For instance, modular invariance of the closed-string holomorphic partition function imposes more restrictive conditions than anomaly cancellation alone \cite{Candelas:1985en,DHVW,Angelantonj:2002ct, SW,Choi:2017luj,Sethi:1996es,Duff:1996rs}. Furthermore, string theory involves high rank tensor fields which induce more general anomaly cancellation mechanisms such as the Green--Schwarz mechanism. These provide important guidelines for evading the swampland \cite{Vafa:2005ui}.

Although the consistency conditions allow for many different vacuum configurations, we have also seen that many of them are dynamically connected. In other words, spontaneous symmetry breaking relates different vacua and the corresponding moduli spaces turn out to be linked. Brane separation and recombination are good examples \cite{Cremades:2002cs, GomezReino:2002fs, Hashimoto:2003pu, Douglas:2004yv, Unif, Kumar:2006yg}. Dual to this process is a phase transition between small instantons and heterotic 5-branes \cite{Witten:1995gx}: by emitting and absorbing branes, the structure group of the instantons changes, and hence the gauge symmetry and matter spectrum can also change.  Increasingly many phenomena have been unified under such dualities.  For instance, recently it has been understood that twisted strings localized on orbifolds can take part in the transition \cite{Choi:2019ovk}. 

Protected by supersymmetry and BPS conditions, these dynamical transitions are continuous and energy-cost free.  
In general these mechanisms change the local chirality and reorganize the spectrum, while the overall chirality is preserved and counted by the global consistency conditions. This may explain the origin of chirality and the absence of anomalies in the Standard Model. 

For every gauge symmetry breaking source, we may consider such a dynamical transition.
It is then natural to ask whether there exists a similar dynamical transition in F-theory. In this work we show that there is a phase transition between M2/D3-branes and $G$-instantons. A $G$-instanton is defined as a classical solution to the self-dual equation of the four-form field strength $G$ on a four-complex-dimensional manifold. In M-theory there is a three-form tensor field $C$ with corresponding field strength
\begin{equation} \label{GFlux}
  G=dC \, .
\end{equation}
In what follows, we borrow this M-theoretical description and apply it in the $T$-dual theory of F-theory.
In the F-theory context, the $G$-instanton may shrink to zero size and undergo a phase transition into a D3-brane. The total chirality is preserved along with the global consistency condition \cite{Sethi:1996es,Duff:1996rs}
\be \label{FGC}
 \frac{\chi(Y)}{24} =  \frac{1}{8 \pi^2} \int_{Y}  G \wedge G + n \, ,
\ee
where $\chi(Y)$ is the Euler characteristic of the Calabi--Yau fourfold $Y$ on which F-theory is compactified, and $n$ is the number of D3-branes filling the noncompact dimensions.  Each term in \eqref{FGC}, especially $n$, must be non-negative integer to ensure the absence of anti-branes and preserve to ${\cal N}=2$ or  ${\cal N}=1$ supersymmetry in four dimensions.

We may write the condition (\ref{FGC}) in a more democratic form, using $\chi(Y) =  \int_{Y} c_4(Y),$ as
\be \label{FGCdiff}
 0 =   d*dC =  \frac{1}{24} c_4(Y) - \frac{1}{8 \pi^2} G \wedge G - \sum_{a=1}^n \delta^{(8)} (y-y_a) \, ,
\ee
where the interpretation in terms of $d * dC$ is possible in eleven and twelve dimensions \cite{twelveaction}.
We see that the transition is natural because the delta function can be interpreted as either (i) a D3-brane source for the equation of motion for $C$ or (ii) a part of the quantized $G$-flux.
 
We first show that this phase transition is dual to that between small instantons and 5-branes  in the heterotic string \cite{Witten:1995gx}. 
It is known \cite{Aspinwall:1994rg} that the phase transition and emergence of a horizontal brane corresponds to blowing-up in the base of F-theory. However this transition may also occur in the case of a vertical brane. As such, it should be translated to a phase transition of $G$-instantons.  Therefore we may prove that there is a transition between small $G$-instantons and D3-branes.
 We can understand their quantitative properties by group-theoretical invariants. 

$G$-flux is an important source for gauge symmetry breaking as well as chirality.  It is induced to the branes and hence to matter curves giving rise to a chiral spectrum in four dimensions \cite{Tatar:2006dc,Hayashi:2008ba}.
An important application is to break the Grand Unification group down to the Standard Model group using a line bundle flux along the hypercharge direction, which can give different chirality for doublet and triplet Higgses \cite{Beasley:2008kw,Donagi:2008kj,Blumenhagen:2008aw,Marsano:2009gv}. Since the phase transition of small $G$-instantons may change the gauge symmetry breaking direction, we may apply it to a phenomenological model. To this end, we consider the toy example of an $SU(5)$ gauge group dynamically broken down to its subgroup by a $G$-instanton phase transition. 

Although D3-branes do not directly contribute to anomaly cancellation in four dimensions \cite{Cvetic:2012xn}, they can contribute indirectly by converting to a $G$-instanton. We demonstrate chirality change in our toy examples and verify that anomaly cancellation is preserved, both before and after the phase transition.

\section{Gauge theory from duality}

We first review the duality between F-theory and heterotic string theory. We establish quantitative relations between the F-theory geometry and the heterotic gauge field. Following this we identify the background gauge bundle.

\subsection{Gauge fields from the three-form field}

We use the duality between F-theory on K3 and the heterotic string on $T^2$ to show the relation between the M-theory three form $C$ and the gauge field $A$ of the heterotic string.
We may expand the field strength $G$ along the harmonic $(1,1)$-forms of the K3 surface \cite{Dasgupta:1999ss},
\be \label{GExp}
 \frac{G}{2\pi} = \sum_I F^I \wedge e_I\, , \quad  e_I  \in H^{1,1}(\K) \, .
\ee
This explains how the Yang--Mills field strength $F$ can acquire non-Abelian structure from an Abelian field $G$.

Nonabelian structure arises if the K3 develops a singularity by shrinking some cycle, which determines the unbroken set of simple gauge groups $\g$. Then we have
\be \label{GGExp} \begin{split}
 \frac12 \int_\K \frac{G}{2\pi} \wedge \frac{G}{2\pi} &= 
 \frac12  \sum_{I,J=1}^{20} \int_{\K} e_I \wedge e_J \wedge F^I \wedge F^J  \\
  &= - \frac12 \sum_{i,j=1}^{8} A_{E_8}^{ij} F_1^i \wedge F_1^j - \frac12 \sum_{i,j=1}^{8} A_{E_8}^{ij} F_2^i \wedge  F_2^j + \frac12 \sum_{a=1}^2 \sum_{i,j} {U}^{ij} F^i_a  \wedge F^j_a \, ,
 \end{split}
\ee
where $U=(\begin{smallmatrix} 0&1\\1&0 \end{smallmatrix})$. 
For each simple Lie algebra component $\g$, the McKay correspondence holds between the intersection numbers and the symmetrized Cartan matrix of the algebra $\g$, 
\be \label{QuadForm}
 \frac{2}{(\alpha^i , \alpha^i)} A_{\g}^{ij} \equiv ( \alpha_{\g}^{(i)\vee} , \alpha_{\g}^{(j)\vee}) =  - \int_{\K} e_i \wedge e_j\, ,
\ee
where the indices $I,J$ are appropriately arranged in (\ref{GGExp}) and $( \cdot ,\cdot )$ is the inner product of root vectors $\alpha^{(i)}$  and/or coroot vectors $\alpha^{(i)\vee}$. Here we use the Chevalley basis, which has the commutation relations  \cite{Fuchs:1997jv}
\begin{align*}
 &[H^i, H^j] = 0 \, , \\
 &[H^{i}, E^j_\pm] = \pm A^{ji} E_\pm^j \, ,\\
 &[E^i_+,E^j_-] = \delta_{ij} H^j \, , \\
 &({\rm ad}_{E^i_\pm})^{1-A^{ji}} E^j_\pm = 0 \, ,
 \end{align*}
where we have suppressed the algebra index. Each set of Cartan generators $H^i$ and ladder operators $E^i_\pm$ is related to a simple root $\alpha^{(i)}$, and we define ${\rm ad}_A B \equiv [A,B]$. Also in the second and fourth lines the signs are correlated. These matrices are represented in the fundamental basis, which satisfies 
\be \label{TraceFund}
 \tr T^a T^b = \frac{(\theta,\theta)}{2} \delta^{ab} \equiv \kappa^{ab} \, . 
\ee
This also defines the Killing form $\kappa^{ab}$ serving as the metric on the algebra space.
Here $\theta$ is the highest weight and we employ the convention $(\theta,\theta)=2$.

Ultimately we are dealing with the real Lie algebra: in the Chevalley basis, the diagonal element $2=A_{SL(n,{\mathbb R})}^{ii}$ corresponds to the normal form of the algebra $SL(n,{\mathbb R})$, for instance \cite{Fuchs:1997jv}. If the above $T^a$ generates $SL(n,{\mathbb R})$, the $SU(n)$ is generated by $i T^a$ having the diagonal element $ -2 = -A_{SU(n)}^{ii}$. We may extend this convention to the exceptional groups.

The killing form is proportional to other traces in different representations. In particular, for the adjoint representation, the normalization constant is the dual Coxeter number $h^\vee_{\g}$,
$$
 \Tr T^a T^b 
 = h^\vee_{\g} (\theta, \theta) \delta^{ab} = 2 h^\vee_{\g} \kappa^{ab} \, , 
$$ 
where $\Tr$ means the trace over the adjoint matrices.  In the Chevalley basis this decomposes into \cite{Fuchs:1997jv}
\begin{align}
\frac{1}{h^\vee_{\g}}    \Tr  H^{i} H^{j} &=   {\red -}\frac{2}{(\alpha^i , \alpha^i)} A^{ij} = \kappa^{ij} \, , \label{Killing1} \\
\frac{1}{h^\vee_{\g}}   \Tr E^\alpha E^\beta& ={\red -}\frac{2}{(\alpha, \alpha)} \delta_{\alpha,-\beta} = \kappa^{\alpha,-\beta} \, , \label{Killing2} \\
\frac{1}{h^\vee_{\g}}   \Tr E^\alpha H^i &=  0 = \kappa^{\alpha i} \, . \label{Killing3}
\end{align}

Since the instanton number is the generalization of the second Chern class, which is an integer in the fundamental representation, we need the normalization 
\be \label{Reduction}
\begin{split}
\frac{1}{2 h^\vee_{\g}}  \Tr F \wedge F &= \frac{1}{2 h^\vee_{\g}}  \sum_{a,b = 1}^{\text{dim\,} \g}\Tr T^a T^b F^a \wedge F^b = \frac{1}{2 h^\vee_{\g}} \sum_{a,b = 1}^{\text{dim\,} \g} \kappa^{ab} F^a \wedge F^b  \\
 &=\frac{1}{2 h^\vee_{\g}}  \sum_{i,j=1}^r \Tr H^i H^j F^i \wedge  F^j + \frac{1}{2 h^\vee_{\g}}  \sum_{\alpha,\beta \in \Phi_{\g}} \Tr E^\alpha E^\beta F^\alpha \wedge  F^\beta \\
  &= {\red -}\frac12  \sum_{i,j} A_{\g}^{ij} F^i \wedge F^j {\red -} \frac12 \sum_{\alpha \in \Phi_{\g}}  F^{\alpha} \wedge F^{-\alpha} \, .
\end{split}
\ee 
Here $\Phi_{\g}$ is the set of the root vectors and we used the Killing form (\ref{Killing1})-(\ref{Killing3}). 
Since $E_8$ and its subgroups embedded therein are self-dual, we fix $(\alpha, \alpha)=2$ so that the normalization
yields integer instanton numbers upon integration over a four-cycle.
Thus (\ref{GGExp}) contains the correct contents and normalization for the Cartan components.

It is known \cite{Weigand:2018rez} that the off-diagonal parts describing $W$-bosons are sourced by the D3-branes wrapped on the two-cycles $e^i$ of \eqref{GExp} with the wrapping number forming the weight vectors in the Dynkin basis.

\subsection{Background vector bundles}

In F-theory, the gauge symmetry is configured by the singular ellptic fiber by choosing the elliptic Calabi--Yau fourfold $\pi: Y \to B'$.  The setup is described by the Weierstrass equation,
\be \label{Weierstrass}
 y^2=x^3+fx+g \, ,
\ee 
with $f$ and $g$ being sections of powers of the canonical bundle of $B'$. Its discriminant locus $D: \{ 4f^3+27g^2=0\}$ determines the cycle supporting the worldvolume gauge theory. The unbroken gauge group $\h$ can be identified by the Kodaira--Tate algorithm \cite{Bershadsky:1996nh}. We leave the details of the construction to review papers, e.g. \cite{Weigand:2018rez}.

We define the Euler characteristic of a singular Calabi--Yau as that of the resolved geometry. As seen in (\ref{GGExp}), the maximal unbroken gauge symmetry is $E_8 \times E_8$. (We also have additional $U(1)$ factors, which we neglect from now on.) The corresponding Calabi--Yau manifold $Y_{\text{sing}}$ is also the most singular, with Euler characteristic \cite{Andreas:1999ng,Esole:2019ynq,Braun:2018ovc}
\be \label{SingularEuler}
 \frac{\chi (Y_{\text{sing}})}{24} = \int_{B} (c_2 + 11 c_1^2 ) 
 \, .
\ee
Here, using the elliptic fibration structure and the vanishing of the first Chern class, we may express the Euler characteristic using Chern classes of the base $B'$ of the elliptic fibration \cite{Sethi:1996es,Klemm:1996ts}. Then the duality between F-theory and the heterotic string requires that the base $B'$ is a $\P^1$ fibration over a two-base $B$, $\pi'': B' \to B$. The fiber is the projectivization of $\P(\O_{B} \oplus t)$, where $t$ is the normal bundle to $B$ in $B'$. Here and in what follows we understand Chern classes without an argument as those of $B$, $c_k = c_k(B)$. It is known that the Euler characteristic does not depend on the method of resolution, that is, does not depend on the choice of flop.

A smoother singularity may be induced by deformation and we obtain a smaller unbroken gauge group. This is done by adding lower-order terms in the Weierstrass equation. 
This deformation will give an additional contribution to the Euler characteristic --- it is known that this translates precisely to the second Chern class of the vector bundle on the heterotic side, i.e.
the correction to the Euler characteristic,
\be
 \chi(Y_{\rm tot}) = \chi (Y_{\text{sing}}) + \Delta \chi(Y_\g) \, ,
\ee
is of the form \cite{Andreas:1999ng,Esole:2019ynq,Marsano:2011hv,Blumenhagen:2009yv}
\be
 \Delta \chi(Y_\g) = \int_B \tilde e(\g) \, ,
\ee
where 
\be \label{GeneralEuler}
 \tilde e(\g) \equiv  h^\vee_{\g} (n_\g \eta (\eta - (\rk \g+1) c_1) +  \dim \g \, c_1^2 ) \, .
\ee
Here $\eta$ is a combination of $c_1$ and $t$, which is to be determined shortly. Also $n_\g$ is the instanton number corresponding to the non-Higgs{\red a}ble cluster.  We have $n_\g = 3,4,6,8,12$ for $SU(n), SO(n), E_6, E_7, E_8$, respectively.
Note that although it is described by group-theoretical quantities such as the dimension $\dim \g$, rank $\rk \g$ and dual Coxeter number $h^\vee_\g$, this contribution is purely geometric.  We do not need to assume the duality to the heterotic string \cite{Friedman:1997yq,Andreas:1999ng,Esole:2019ynq,Marsano:2011hv}.

Nevertheless, we should have the same unbroken gauge group $\h$ {\red on} the heterotic side.
In particular, we consider the dual $E_8\times E_8$ heterotic string on an elliptic Calabi--Yau threefold $X$ which is a fibration over the same two-base $B$, $\pi_{\rm h}: X \to B$. 
The unbroken gauge group $\h$ is the commutant to the structure group $\g$ of the vector bundles $V_1$ and $V_2$ in $E_8\times E_8$. Each of them gives an additional contribution to the second Chern class,
\be \label{SecondChern}
 c_2(V) = \eta \sigma - e(\g) \, ,
\ee
where $\sigma$ is the section of the base $B$ in $X$.

For example, if we wish to have an unbroken $SU(5)$ gauge group, we need background gauge bundles $V_1$ and $V_2$ whose structure groups are the commutant to $SU(5)$ and $E_8$, respectively, such that \cite{Friedman:1997yq,Andreas:1997ce}
\begin{align}
 c_2(V_1) &= \eta_1 \sigma -  5 c_1^2 - \frac58 \eta_1 (\eta_1 - 5 c_1) \, , \\
 c_2(V_2) &= \eta_2 \sigma- 310 c_1^2  - 15  \eta_2 (\eta_2 - 9 c_1) \,.
\end{align}
Meanwhile the second Chern class for the tangent bundle of $X$ is \cite{Friedman:1997yq}
\be
 c_2(X) = c_2 + 11 c_1^2 + 12 \sigma c_1 \, .
\ee
Note that it is the same as the integrand (\ref{SingularEuler}) plus $12 \sigma c_1$. 
We may have a mismatch between them \cite{Friedman:1997yq},
\be \label{BianchiChern}
 c_2(X) - c_2(V_1) - c_2(V_2) \ne 0 \, ,
\ee
apparently failing to satisfy the global consistency condition, namely the Bianchi identity for the Kalb--Ramond field strength $H$,
\be \label{HetGlobalCons}
 \begin{split}
 0 = \frac{2}{\alpha'}dH
 & = \frac{1}{2} \tr R \wedge R - \frac{1}{ 2 h^\vee_{E_8}} \Tr F_1 \wedge F_1 - \frac{1}{ 2 h^\vee_{E_8}}  \Tr F_2 \wedge F_2 \\
 &= c_2 (X) - c_2(V_1) - c_2(V_2)   \, ,
\end{split}
\ee
where $\tr$ is the trace over the vector representation of the Lorentz group while $\Tr$ is taken over the adjoint representation of $E_8$. The dual Coxeter number of $E_8$ is $h^\vee_{E_8}=30$.
This mismatch should be accounted for by heterotic 5-branes, providing magnetic sources for the Kalb--Ramond field $B$.

\subsection{$\bm G$-flux}

The gauge symmetry from the singular fiber can be broken further by $G$-flux (\ref{GFlux}). While there can be nontrivial flux that preserves the gauge symmetry, in any case it will induce a chiral spectrum in four dimensions. The gauge bosons of the broken symmetry acquire mass via the St\"ukelberg mechanism \cite{Donagi:2008kj}. In addition, this background flux induces the D3-charge as in (\ref{FGC}) and (\ref{FGCdiff}).

On the heterotic side, the $G$-flux provides an extra component of the vector bundle and thus also breaks the gauge group. This is reflected in the second Chern class (\ref{SecondChern}), 
\be 
 c_2(V) = \eta \sigma - e (\g) + \frac{1}{8 \pi^2} \int_\K G \wedge G \, .
\ee
The relation between M-theory and $E_8\times E_8$ heterotic string theory  imposes the Freed--Witten quantization condition for the cohomology of $G$ \cite{Witten:1996md},
\be \label{Quantization}
 \frac{G}{2\pi} + \frac12 c_2( Y) \in H^4(Y, \Z) \, .
\ee
That is, if $\frac12 c_2 (Y)$ is not integrally quantized, then $G$ should be also half-integrally quantized.

Supersymmetry restricts $G$ to be a $(2,2)$-form. This flux should not break the Poincar\'e symmetry on the F-theory side, thus one of the leg of four-form field $G$ should be along the fiber. This is expressed as \cite{Dasgupta:1999ss}
\begin{align}
 G & \cdot Z \cdot D_a = 0 \, ,  \label{Zvertical} \\
 G & \cdot D_a \cdot  D_b = 0 \, , \label{vertical}
\end{align}
where $Z$ is the section of the elliptic fibration and $D_a, D_b$ are pullbacks of base divisors in $B'$. Here and in what follows, a dot product among divisors means intersection inside the Calabi--Yau manifold $Y$.

When the fiber becomes singular, giving rise to the gauge group $\h$, we may blow-up these singularities to obtain exceptional divisors $E_i$. It turns out that these are $\P^1$ fibrations over the discriminant locus $D$, whose fibers are the above $e_i$, c.f. \eqref{GExp}. Again, these obey a generalized McKay relation, that is, they are in one-to-one correspondence with the roots of the Lie algebra,
\be \label{McKay}
 E_i \cdot E_j \cdot D_a \cdot D^b = - A_{ij} \delta_a^b \, ,
\ee
where $A_{ij}$ is the Cartan matrix of $\h$.

It is known \cite{Weigand:2018rez} that the vertical divisor can be made of a bi-product of two (1,1)-forms. So we consider fluxes of the type
\be
 E_i \wedge D_a \, .
 \ee
Here we assume that there are no $U(1)$ gauge groups arising from rational sections. We may also construct $G$-flux dual to a matter curve having the form
\be
 \quad  E_i \wedge E_j \, ,
\ee
which should also satisfy the conditions (\ref{Zvertical}) and (\ref{vertical}). Furthermore, a pair of $E_{i}$ divisors should be chosen which do not produce a nonzero value upon integration along the K3.

Usually we construct a $G$-flux preserving the whole gauge symmetry arising from the singularity, 
\be \label{CartanFlux}
 G \cdot E_i \cdot D_a  = 0 \,, \quad \alpha^i \in \Phi_{\rm s}(\h) \, ,
\ee
in order to obtain a chiral spectrum.  However, we may also relax one or more of these conditions in order to introduce additional symmetry-breaking $G$-flux.

\subsection{Chirality}

An important role of $G$-flux is to induce chirality in four dimensions. 

For localized matter fields on the matter curve $\Sigma_{\bf R}$, the local gauge symmetry on the discriminant locus is enhanced, and the branching of the adjoint of the enhanced gauge group determines the representation $\bf R$ \cite{Hayashi:2008ba}. The curve is thus promoted to a matter surface, a $\P^1$ fibration over the matter curve \cite{Krause:2011xj}. To be precise, the fiber is a linear combination of $\P^1$s  reflecting the weights of components of $\bf R$. Thus we have as many slices of matter surfaces as the dimension of $\bf R$. We define the matter surface ${\cal S}_{\bf R}$ as that corresponding to the highest weight component of $\bf R$.

A chiral fermion in six dimensions compactified on a smooth two-manifold, which is a matter curve $\Sigma_{\bf R}$ in our case, gives vector-like fermions in four dimensions. However the zero modes may become chiral if we turn on nontrivial magnetic flux on $\Sigma_{\bf R}$.  The $G$-flux induces magnetic flux on the matter curve, which is derived from the that induced on the matter surface ${\cal S}_{\bf R}$ \cite{Krause:2011xj}{\red,}
\be
 \chi({\bf R}) = \int_{{\cal S}_{\bf R}} G \, .
\ee

If the $G$-flux breaks the gauge symmetry given by the singularity, it gives rise to different chiralities in the different branched representations. The gauge symmetry is broken to $\g \to \h \times \h'$, where fluxes $F^I$ belong to the structure group $\h'$. Accordingly, the representation also branches as $\bf R \to (r,1)+(1,r')$. The latter two differ by their respective roots, and the nontrivial intersection with the $G$-flux may induce different chiralities in each representation. In particular, if the $G$-flux is proportional to a Cartan element $E_I$, then the resulting chirality is proportional to the $U(1)$ charge along the $E_I$ direction.

In addition, the vector multiplet branches and its `off-diagonal' components $\bf R$ and $\bf \bar R$ may become chiral. The chirality is determined by the Riemann--Roch--Hirzebruch index theorem  \cite{Beasley:2008dc},
\be \label{RRH}
 \chi({\bf R}) \equiv n_{\bf R} - n_{\bf \bar R} =  \int_B c_1(F) \wedge c_1(B) \, .
\ee
It is defined as the number of zero modes for a complex representation ${\bf R}$ minus that of ${\bf \bar R}$.

\section{Small $\bm G$-instantons}

Using heterotic/F-theory duality, we show that small $G$-instantons may shrink and become D3-branes via a phase transition. F-theory on a Calabi--Yau threefold is dual to heterotic string theory on a K3 manifold. This is extended in higher dimensional Calabi--Yau manifolds by replacing the common $\P^1$ base with any rational manifold $B$. 
For each base divisor $C_a$ of $B$, we may define Poincar\'e-dual curves $C^b$ satisfying 
\be \label{BBasis}
 \int_B C_a \wedge C^b =  \delta_a^b \, .
\ee
Sections $\sigma C_a$ are curves in $X$, while pullbacks $\pi^*_{\rm h} C^a$ are surfaces in $X$ providing Hodge  duals of $\sigma C_a$, which arises from the relation (\ref{BBasis}) since
\be \label{XIntersecs}
  \delta^a_b = \int_X \sigma B \wedge \pi^*_{\rm h}  C^a \wedge \pi^*_{\rm h}  C_b = \int_X \sigma C^a  \wedge \pi^*_{\rm h} C_b \, ,
\ee
where we have implicitly used the essential property of the fiber $E$ that
\be
 \int_X \sigma B \wedge E = 1 \, .
\ee

\subsection{Small-instanton transition}

There is a transition between small instantons and 5-branes \cite{Witten:1995gx}.  If an instanton shrinks to zero size, the gauge symmetry is recovered and the instanton becomes a 5-brane. The zero size instanton is described by a delta function, which also describes a 5-brane source for the equation of motion of $*H$:
\be
 0 = \frac{2}{\alpha'}dH =  \frac{1}{2} \tr R \wedge R
 - \frac{1}{2 h^\vee_{E_8}} \Tr F_1' \wedge F_1'  - \frac{1}{2 h^\vee_{E_8}}  \Tr F_2' \wedge F_2 '  
  -  \sum_{A=1}^{n} \delta^{(4)}(C_A) \, .
\ee
Here each delta function is nonvanishing along the locus of a curve $C_A$. We may think of the argument of the function as being the zeroes of the defining equation of $C_A$. Being protected by ${\cal N}=2$ supersymmetry, the transition takes place without energy loss. 

Depending on the direction of $F \wedge F$, different 5-branes may or may not wrap the elliptic fiber $E$. If a 5-brane wraps the elliptic fiber of the heterotic string, we refer to it as a vertical brane; if the a brane is transverse to $E$, we call it a horizontal brane.

The properties of instantons differ depending on the type of string.
For instance, the $SO(32)$ heterotic string theory is dual to the type I string, and the gauge group is constructed via D9-branes/O9-planes. Meanwhile the heterotic 5-brane is dual to a D5-brane. It is well known that the D9-D5 system describes instantons and the orientifold merely introduces a further projection on the Chan--Paton factor. The finite-size instantons correspond to flux sourced by D5-branes while shirinking instantons can become an unbounded state under a phase transition with no energy cost. Each D5-brane on top of O9-planes describes an $Sp(1) \simeq SU(2)$ gauge group, which is further enhanced to $Sp(k)$ for $k$ coincident branes. Therefore the gauge anomaly is constrained. In the Coulomb branch we may detach the D5-branes, whose locations relative to the D9-branes are described by adjoint-valued scalars in the vector multiplet.

In the $E_8 \times E_8$ heterotic string, a new interval opens up in the strong coupling limit \cite{HW}. It is dual to M-theory compactified on this interval, at whose boundaries exist M9-branes each describing one $E_8$ factor. The resulting 5-brane is an M5-brane and is emitted into the eleven-dimensional bulk. This corresponds to a tensor branch for which we have a new antisymmetric tensor field, whose scalar partner in the tensor multiplet describes the location of NS5-branes.  From the point of view of the M9 branes we cannot simultaneously observe the gauge group described by the 5-branes.

In F-theory the M9-branes are mapped to a stack of 7-branes of appropriate $(p,q)$ charges. This unifies the above descriptions of 7-branes. It is natural that there may be a shrinking flux becoming a pointlike brane in $Y$. We may also have D3-branes, which are  self-dual under the exchange of $p$ and $q$ charges, detatched from the 7-branes. We will demonstrate this using the duality to the heterotic string.

\subsection{$\bm G$-instanton}

For a four-complex-dimensional manifold $Y$, we may consider the self-dual equation for the $G$-flux \cite{Becker:1996gj,Dasgupta:1999ss},
\be \label{GSelfdual}
 *_{Y} G = G \, .
\ee
We refer to the classical solution of this equation as a $G$-instanton. Using the representation $J = i g_{a \bar b} dz^a \wedge dz^{\bar b}$ in the K\"ahler manifold, the primitivity condition $J \wedge G = 0$ implies
\be
 g^{c \bar d} G_{a \bar b c \bar d} = 0 \, .
\ee

In this work, we are interested in $G$-instantons which are dual to heterotic instantons.
Expanding $G$ in terms of the exceptional divisors as in (\ref{GExp}), the self-duality of $G$ in (\ref{GSelfdual}) translates into the (anti-)self-duality
\be \label{FSelfdual}
 *_B F = \pm F
\ee
in $B$, since $E_i$ are (anti-)self-dual in K3. The K3 manifold has signature (19,3), meaning that there are 19 self-dual and 3 anti-self-dual two forms $E_i$. In particular, all Cartan subalgebra elements of $E_8\times E_8$ belong to the self-dual part. 
The self-dual equation of $F$ translates to the Hermitian Yang--Mills (HYM) equation in the heterotic string \cite{Becker:1996gj}, 
\be
 g^{a \bar b} F_{a \bar b} = 0 \, ,
\ee
along the Cartan directions.  Ten-dimensional supersymmetry imposes this condition on a general gauge field --- the instanton solution is a special case of this.

On the heterotic side there is no difference between the base $B$ and the fiber $E$ directions. Thus we may consider another self-duality of the form
\be
 *_{\pi^*_{\rm h} C^a} F = \pm F \, ,
\ee
where we may take an element $C_a \in H^2(B,\Z)$ such that the pullback of its Poincar\'e dual is a four-cycle, $\pi^*_{\rm h} C^a \in H_4 (X,\Z)$.  The fiber $E$ is not accessible in the F-theory side, so this does not appear as a $G$-instanton.

\subsection{Horizontal brane}

There is a transition between small instantons and 5-branes, which corresponds to blowing-up in the base $B'$ of an elliptic fibration of $Y$ on the F-theory side.
If the components of $F \wedge F$ lie on the elliptic fiber,
\be
 \frac{1}{2 h^\vee_\g}  \Tr F \wedge F |_{\pi^*_{\rm h} C^a} =  \frac{1}{2 h^\vee_{\g'}} \Tr F' \wedge F'  |_{\pi^*_{\rm h} C^a} + \delta(\sigma C^a) \, ,
\ee
then the corresponding instanton shrinks to a point and the resulting 5-brane becomes transverse to the elliptic fiber. This is known as a horizontal brane and wraps a four-cycle $\pi^*_{\rm h} C^a$. Recall that in our notation, $C_a$ is the section of a curve in the homology, and it is Poincar\'e dual to a four form. From (\ref{XIntersecs}) it is natural to denote
\be
 \delta^{(4)}(\sigma C^a) = \sigma C^a.
\ee

It is known \cite{Aspinwall:1994rg} that the horizontal instantons are `conformal matter' localized at the intersections between discriminant locus components, when the base manifold is the $\P^1$ fiber bundle $\pi'' : B' \to B$.
The discriminant lies on the class $D$, given by adjunction as
$$
 D = 12 c_1(B') = 12 c_1 + 24 r + 12 t \, ,
$$
where $r$ is the section of the above fibration $\pi''$.
Since we have two order ten singularities II$^*$ for $E_8$ at $r$ and $r+t$, respectively, we may decompose the discriminant locus into irreducible components. The remaining one becomes a horizontal `instanton locus' \cite{Aspinwall:1994rg} 
$$
 D'_{\text{hor}} = D - 10 r - 10 (r+t) = 12 c_1 + 4 r + 2 t \, .
$$
That is, we have $E_8$ instantons when this surface intersects the discriminant locus component at $r=0$ and $r=\infty$, respectively,
\be
 D'_{\text{hor}}|_r = (12 c_1  + 2 r)r = 2 (6c_1 - t) r, \quad D'_{\text{hor}}|_{r+t} = 2 (6c_1+t)(r+t) \, ,
\ee
where we have used the property $r(r +t) =0$. This hypersurface is six dimensional and mapped to a cycle on which heterotic 5-branes are wrapped \cite{Aspinwall:1994rg,Choi:2017vtd}.  The branes are horizontal. 
The extra factor 2 reflects the fact that the singularity is of Kodaira type II: the discriminant is the double zero of the corresponding coefficient of the Weierstrass equation, \textit{i.e.} $\Delta \sim (g_{6c_1 +t} z^5)^2$ \cite{Aspinwall:1994rg}. 
From this, we identify 
\be \label{GenInstNo}
 \textstyle  \eta_1 = \pi^{\prime \prime}_* ( \frac12 D'_{\text{hor}}|_r ) =  6c_1 -t\, , \qquad \eta_2 = \pi^{\prime \prime}_* ( \frac12 D'_{\text{hor}}|_{r+t}) = 6c_1 + t \, ,
\ee
where $\pi^{\prime \prime} (r) =1$. Deformation of the singularity corresponds to the addition of more terms in $f$ and $g$ in (\ref{Weierstrass}), however the coefficient of $ g_{6c_1 +t} z^5$ in $g$, and hence the number of embedded instantons, remains the same.

For the above choice (\ref{GenInstNo}), we do not need horizontal 5-branes because the coefficients of $\sigma$ in $c_2(X)$ and $c_2(V_1)+c_2(V_2)$ are the same, that is, all instantons are contained in $V_1$ and $V_2$. By a phase transition some of the instantons can be emitted as horizontal branes. The above process can be rephrased as
\be \label{HorizontalH5Cond}
\begin{split}
	\eta \sigma & = (6  c_1- t) \sigma =  \frac{1}{2 h^\vee_\g} \Tr F \wedge F \\
	& =  \frac{1}{2 h^\vee_{\g'}} \Tr F' \wedge F'+  C^a  \sigma = \eta' \sigma + C^a \sigma \, .
\end{split}
\ee
In this case, the duality is well-known \cite{Aspinwall:1994rg, MV,Andreas:1999ng,Andreas:1999zv}.

\section{Gauge symmetry breaking via $\bm G$-instanton transitions}

Finally, we study the transition between $G$-instantons and D3-branes using the dual transition between heteortic instantons and vertical 5-branes. In the process we break the gauge symmetry dynamically.  A quantative understanding of this process comes from group theory.

\subsection{Vertical brane}

In the previous section we have seen that there is a transition between small instantons and horizontal 5-branes. From the viewpoint of the heterotic string, there is no difference between the horizontal and vertical directions.  That is, there should also be small instanton transitions {\em along the base directions}.  The component of $F \wedge F |_B$ transverse to the elliptic fiber $E$ shrinks to a point and undergoes a phase transition, 
\be
\frac{1}{2 h^\vee_{\g}} \Tr F \wedge F|_B = \frac{1}{2 h^\vee_{\g'}} \Tr F' \wedge F'|_B + \delta^{(4)} (E) \, .
\ee
The resulting 5-brane is a vertical brane that wraps the fiber $E$.

From the orthonormality of the basis (\ref{BBasis}), a four-form becomes a Dirac delta function supporting $E$,
\be
 \delta^{(4)} (E) = C_a  \wedge C^a \, . 
\ee
Its location in $B$ is a point $a$ at the intersection of two curves, but we are interested in its homology.  There are distinct delta functions associated with each basis curve $C_a$. 
If the curve $C_a$ is is self-dual like the self-dual field $F$,
\be
 C^a = *_B C^a = C_a \, , 
\ee
then the above phase transition can be explained by a transition of the form
\be \label{VerticalFlux}
 \frac{G}{2\pi} = \frac{G'}{2\pi} + P_a \wedge C_a \, .
\ee
Here $P_a$ is a linear combination of exceptional divisors $E_i$ related to the simple roots $\alpha^{(i)} \in \Phi_{\rm s} (\h)$ of the unbroken gauge group $\h$,
\be \label{HetQuant}
 P_a = \sum_i c_a^i E_i, \quad c_a^i \in \Z \, .
\ee
The coefficients are integer in general and determined by the quantization condition (\ref{Quantization}). 
As long as we choose $E_i$ orthogonal {\red to} the structure group of $G$ such that  
\be \label{Gcond}
 \forall c_a^i \ne 0 \, , \quad \alpha^{(i)} \in \Phi_{\rm s} (\h) \quad \Longrightarrow \quad G' \cdot E_i \cdot D_a = 0 \, ,
\ee
we have no cross-term in the expansion
\be
 \frac1{8 \pi^2} G \wedge G = \frac1{8 \pi^2} G' \wedge G' + \frac12 \pi^* C_a \wedge \pi^* C_a \wedge P_a \wedge P_a \, .
\ee
This $P_a$ has one-to-one correspondence with the weight vector $c_a = (c_a^i)$.

The delta function describes a heterotic 5-brane wrapped on $T^2$, where the wrapped volume has been factored out.  It is mapped to a D3-brane on the F-theory side,
\be
 \delta^{(8)}(x-x_a) = \frac12 P_a \wedge P_a\wedge \pi^* C_a \wedge \pi^* C_a \, .
\ee
The normalization factor $1/2$ is set by the relation (\ref{GGExp}).
Some part of the $G$-flux may be emitted and become D3-branes, described by the delta function in (\ref{FGCdiff}).
During the transition the total D3 charge is conserved, so the relation (\ref{FGCdiff}) remains satisfied.

We can see that the number of heterotic 5-branes is always an integer.
This corresponds to the diagonal elements of the Cartan matrix, which denote the Cartan subalgebra.
Since we obtained $e_i$ and $E_i$ from the $H^2$ basis of the K3, For $E_8$ and its subgroup, the number of instantons is guaranteed to be an integer because the product
\begin{align}
- \frac12 \int_\K P_a \wedge P_a &=- \frac12 \int_\K \left( \sum_{i} (c_a^i)^2 E_i \wedge E_i +2 \sum_{i<j} c_a^i c_a^j E_i \wedge E_j \right) \nonumber \\
 &= \sum_{i} (c_a^i)^2 - \sum_{i<j} c_a^i c_a^j A_{ij} \, 
\end{align}
is always an integer if $c_a^i$ are integrally quantized as in (\ref{HetQuant}). 
Hence we see that the instantonic $G$-flux is integrally quantized, so the number of D3-branes should be an integer. Thus the quantitative aspect of the small-instanton transition is guided by the fact that the $E_8$ lattice is even and the $E_8$ algebra is encoded in the intersection structure of K3 through the expansion (\ref{GGExp}).

\subsection{$\bm G$-instantons breaking gauge symmetry}

We wish to use the $G$-instanton transition, which comes from the phase transition of D3-branes, to explain dynamical transitions among vacua.
We may break the gauge symmetry using $G$ by relaxing some of the conditions \eqref{Gcond} \cite{Marsano:2011hv,Braun:2018ovc,Choi:2010nf}.  Let $G^{\h}$ the $\h$-preserving flux consistent with the quantization condition. We have
\be
 G = G^{\h} + G^{\text{inst}} \, ,
\ee
where
\be \label{Ginst}
 \frac{G^{\text{inst}}}{2 \pi}  \equiv \sum_i a_i \Lambda_{(i)} \wedge {\cal F} \, .
\ee
The two-form $\cal F$ satisfies the anti-self-duality condition in $B$,
\be
 *_B {\cal F} = - {\cal F} \, .
\ee
Among the two sign possibilities in (\ref{FSelfdual}), we choose the minus sign in order to ensure that the resulting D3-brane energy density is positive definite.
This is the same form as (\ref{VerticalFlux}) because the first factor $\Lambda_{(
i)}$ is expressed in terms of the exceptional divisors $E_i$, while the second factor corresponds to base divisors. 
One may verify that this $G^{\text{inst}}$ satisfies the conditions (\ref{Zvertical}) and (\ref{vertical}).
The coefficient is fixed because both $G$ and $G^{\h}$ satisfy the quantization condition (\ref{Quantization}).

Since the divisor $P_a$ of \eqref{HetQuant} is expanded in terms of the cycles $E_i$ which satisfy the McKay correspondence with the roots $\alpha^{(i)}$ of the Lie algebra, we can use this to control the symmetry breaking direction. 
Hence we introduce fundamental weight divisors,
\be \label{FundWghts}
 \Lambda_{(i)}= (A^{-1})^{ij} E_j \, ,
\ee
using the inverse $(A^{-1})^{ij}$ of the Cartan matrix (\ref{QuadForm}). By construction they satisfy the orthonormality condition
\be 
 \Lambda_{(i)} \cdot E_j \cdot D_a \cdot D^b = - \delta^i_j \delta_a^b \, .
\ee
This basis requires a stronger quantization condition for the coefficients $a_i$ \eqref{Ginst} because the inverse Cartan matrix $A^{-1}$ has fractional entries, so a larger multiplicity is needed to ensure that all $E_i$ cycles are wrapped an integer number of times.

The induced D3-brane charge is given by
\be \label{GGtransition}
\begin{split}
 \frac{1}{8 \pi^2} \int_Y G \wedge G & = \frac{1}{8 \pi^2}  (G^{\h} + G^{\text{inst}}) \cdot (G^{\h} + G^{\text{inst}}) \\
&  = \frac{1}{8 \pi^2}  (G^{\h})^2 + \frac1{8 \pi^2} (G^{\text{inst}})^2 \, ,
\end{split}
\ee
Owing to the relation (\ref{CartanFlux}) there is no cross term, since
\be
 G^{\h}  \cdot G^{\text{inst}} = G^{\h}  \cdot \sum a_i \Lambda_{(i)} \cdot {\cal F} = 0 \, ,
\ee
where in the second equality we have used the fact that $\Lambda^{(i)}$ is a linear combnation of exceptional divisors $E_i$.  Hence we see that this provides a good description of the phase transition between $G$-flux and D3-branes.

We now apply the transition between $G$-flux and D3-branes to gauge symmetry breaking. The part $\frac12 G^{\text{inst}} \cdot G^{\text{inst}}$ becomes a delta function and describes the location of the D3-branes after the phase transition. Explicitly we have
\be \begin{split}
 \frac{1}{8 \pi^2} (G^{\text{inst}})^2  & = \frac{1}{2}\left(\sum_i a_i  \Lambda_{(i)} \right)^2 \cdot {\cal F} \cdot  {\cal F} \\
 & = -\frac{1}{2} \sum_{i,j} a_i a_j (A^{-1})^{ij}  \int_B {\cal F} \wedge {\cal F} \, .
 \end{split}
\ee
We see that the number of D3 branes emitted is determined by elements of the inverse Cartan matrix of $\h$.

We also wish to translate the above picture to the heterotic side.  The instanton flux $G^{\text{inst}}$ is of the form \eqref{Ginst} which is a simple wedge product of exceptional and base divisors, so it can simply be expanded in Chevalley basis.  The $\h$-preserving flux is more subtle: it corresponds to shift in the second Chern class of the gauge bundle $c_{2}(V_{1})$. However, this `traceless' component cannot be seen from the base geometry, as it is projected out \cite{Friedman:1997yq}.
Nevertheless we can trace the contribution for the heterotic 5-branes. The change of the heterotic flux is, using the reduction (\ref{Reduction}), 
\be \begin{split}
 \frac{1}{2 h^\vee} \int_B \Tr F^{\text{ inst}} \wedge F^{\text{ inst}} 
 & = {\red -}\frac12 \int_B A^{ij} (a_k (A^{-1})_{ik} {\cal F}) \wedge ( a_l (A^{-1})_{jl} {\cal F}) \\
  &= {\red -}\frac12 \sum_{i,j} a_i a_j (A^{-1})^{ij}    \int_B {\cal F} \wedge {\cal F}\, . 
\end{split}
\ee
This is the number of heterotic 5-branes after the transition. Note that this agrees with direct calculation of the number of D3-branes in the F-theory side.

After emission of the flux, the $G$-flux can become ``smaller,'' and thus the unbroken gauge group becomes smaller. Therefore, many vacua having different gauge groups and spectra can be connected.
For example, starting from $SU(5)$ we may break the gauge group down to its subgroups.  We may summarize this data as weight vectors in the Dynkin basis.

\subsection{Example}

As an example, we construct an $SU(5)$ surface with blowups and consider the phase transition that reduces the unbroken gauge group to $SU(4)$ or $SU(3) \times SU(2)$. The exceptional divisors $E_1,E_2,E_3,E_4$ satisfy the correspondence (\ref{McKay}), whose relations are summarized by the Cartan matrix
$$
 A^{ij} = \begin{pmatrix} 2 & - 1 & 0 & 0 \\ -1 & 2 & -1 & 0 \\  0 & -1 & 2 & -1 \\ 0 & 0 & -1 & 2 \end{pmatrix} \, .
$$ 
We adopt the resolution used in \cite{Krause:2012yh,Marsano:2011hv} where we take $E_2 \cdot E_4$ to be invariant.

The $SU(5)$-preserving flux is obtained from the solution 
\be \label{su5flux}
G^{SU(5)} \cdot E_i \cdot D_a = 0, \qquad i=1,2,3,4 \, , \quad \alpha^{(i)} \in \Phi_{\rm s}(SU(5)) \, .
\ee
These conditions can be imposed on a $G$-flux ansatz satisfying the quantization conditions \eqref{Quantization} and verticality conditions \eqref{Zvertical} and \eqref{vertical}, namely
\begin{equation} \label{G4ansatz}
\frac{G}{2\pi} = \sum_{i=1}^{4} E_{i} \cdot (a_{i}c_{1}(B') + b_{i}B) + \frac{1}{2}(E_{2} + E_{3})\cdot c_{1}(B') + \left(p+\frac{1}{2}\right) E_{2}\cdot E_{4} \, , 
\end{equation}
where $B$ is the section of the two-base in $Y$ and flux quantization restricts $a_{i}, b_{i}, p \in \mathbb{Z}$.  Imposing $SU(5)$-invariance \eqref{su5flux} leads to a unique $G$-flux,
\begin{equation} \label{G4su5}
 G_{\lambda}^{SU(5)} = \lambda\left(E_{2} \cdot E_{4}+\frac{1}{5}(2E_1-E_2+E_3-2E_4) \cdot c_{1}(B')\right) \equiv \lambda\, \hat{G}^{SU(5)} \, ,
\end{equation}
where the quantization condition on $\lambda$ can be expressed as
\begin{equation}
\lambda = -\frac{5}{2}\left(1+2a_{\lambda}\right) \, , \qquad a_{\lambda} \equiv a_{3} - a_{1} \in \Z \, .
\end{equation}

If we relax (\ref{su5flux}) by requiring only three of the four conditions, we may introduce additional flux which spontaneously breaks $SU(5)$ to a subgroup, either $SU(3)\times SU(2)$ or $SU(4)$ depending on the breaking direction.  Relaxing the $i$th condition of \eqref{su5flux} and imposing the remaining constraints on the ansatz \eqref{G4ansatz} yields expressions of the form
\begin{equation} \label{G4breaking} 
\begin{split}
 G_{\textrm{tot}}^{(i)} &= G_{\,\lambda}^{SU(5)} + G^{(i)} \, ;\\
  \frac{G^{(i)}}{2\pi} &= a^{(i)} \left(\Lambda_{(i)}\cdot c_{1}(B') + n^{(i)}\frac{\hat{G}^{SU(5)}}{2\pi} \right) + b^{(i)}\Lambda_{(i)}\cdot B \, ,
\end{split}
\end{equation}
where $a^{(i)}$ and $b^{(i)}$ are new parameters constructed from linear combinations of the $a_{i}$s and $b_{i}$s, respectively, while explicit calculation gives $n^{(i)} = (-2,1,4,2)^{i}$.
Here we have also introduced the divisor related to fundamental weights as in (\ref{FundWghts}), where the inverse Cartan matrix for $SU(5)$ is defined as
\be
 (A^{-1})^{ij} = \frac15 \begin{pmatrix} 4 & 3 & 2 & 1  \\ 3 & 6 & 4 & 2 \\ 2 & 4 & 6 & 3 \\1 & 2& 3 & 4 \end{pmatrix} \, .
\ee

Flux quantization requires $a_{i}, b_{i} \in \Z$, which in turn fixes $b^{(i)}\in 5\Z$.  This can be seen heuristically from the fact that a factor of $5$ is required to cancel the $1/5$ in $\Lambda_{(i)}$ to recover an integral cycle.  However the quantization of $a^{(i)}$ is more subtle due to the $SU(5)$-invariant contribution $\hat{G}^{SU(5)}$.  Quantization of the total flux in practice only requires it to be either an integer or odd half-integer.  Only if we require each term in \eqref{G4breaking} to be quantized separately is $a^{(i)}$ further constrained to be a multiple of five.  If this is the case, the resulting flux can be simplified to the form
\begin{equation}
 \frac{G_{\textrm{tot}}^{(i)}}{2\pi} = \frac{ G_{\lambda'}^{SU(5)}}{2\pi} + 5\Lambda_{(i)}\cdot(ac_{1}(B')+bB) \, ,
\end{equation}
where $a^{(i)} \equiv 5a$, $b^{(i)} \equiv 5b$, and $\lambda'$ is defined such that
\begin{equation}
a_{\lambda'} \equiv a_{\lambda} - n^{(i)}a \, 
\end{equation}
for each $i$.  Note that while $a$ may not always be an integer, the combination $n^{(i)}a$ is always integral.  Finally, if we require $G^{(i)}$ to be self-dual, given that the exceptional divisors $E_{i}$ are self-dual on the K3, the corresponding base divisor on $B$ must be self-dual.  However this is generally not the case for (the pullback of) $c_{1}(B')$, so for an instanton transition we may restrict
\begin{equation} \label{azero}
a = 0 \, .
\end{equation}
Thus $\lambda$ cannot change during an instanton transition.

Finally, note that the ansatz \eqref{G4ansatz} may be too restrictive, since it uses a basis of fluxes specifically designed to cancel $SU(5)$ breaking terms: in the conditions \eqref{su5flux}, double and triple intersections of the $E_{i}$s can always be reduced to terms proportional to $c_{1}(B')$ or $B$ only.  Hence if we relax any of the conditions \eqref{su5flux}, further contributions of the form $c^{ai}E_{i}\cdot D_{a}$ may be allowed, with $D_{a}$ the pullback of base divisors and $c^{ai}\in\Z$.  By the same arguments, and taking into account \eqref{azero}, the most general flux breaking $SU(5)$ along the $i$th Cartan direction takes the form
\begin{equation} \label{G4genbreaking}
 \frac{G^{(i)}}{2\pi} = 5\Lambda_{(i)}\wedge \sum_a c^{a}C_{a} \, ,
\end{equation}
where $c^{a}\in\Z$ and we have expanded the discriminant locus $B$ in the basis $C_{a}$.

To construct a self-dual flux, we require that the base divisors $C_{a}$ are self-dual.  The $C_{a}$ divisors form a self-dual orthonormal basis in the cohomology $H^{2}(B,\Z)$, \textit{i.e.}
\begin{equation}
\int C^{a}\wedge C_{b} = - \delta^{a}_{b} \, ,
\end{equation} 
where $C^{a} = - C_{a}$ for anti-self-dual divisors.  If we consider a single $C_a$ component, the resulting instanton contribution is then given by
\begin{equation}
\frac{1}{8\pi^2}\int G^{(i)}\wedge G^{(i)} = \frac{25}{2}(A^{-1})^{ii} (c^{a})^{2} \, .
\end{equation}

For example, taking the breaking direction as the fourth direction,
\begin{equation} \label{FourthDir}
 \frac{ G^{(4)}}{2\pi} = 5\Lambda_{(4)} \wedge c^{a}D_{a} = (E_1 + 2E_2 + 3 E_3 + 4 E_4) \wedge c^{a}D_{a} \, ,
\end{equation}
the instanton contribution for $SU(4)$ invariance is
\begin{equation}
\frac{1}{8\pi^2}\int G^{(4)}\wedge G^{(4)} = 10 (c^{a})^{2} \, .
\end{equation}
Similarly, if we choose $i = 3$ in order to preserve $SU(3)\times SU(2)$, the instanton becomes
\begin{equation}
\frac{1}{8\pi^2}\int G^{(3)}\wedge G^{(3)} =15 (c^{a})^{2} \, .
\end{equation}
Here we see that $10$ D3-branes are absorbed to break to $SU(4)$, while $15$ are needed to reach $SU(3)\times SU(2)$.

We can generalize this result to the $SU(n)$ case: $G$-flux of the form
\begin{equation}
 \frac{G^{(i)}}{2\pi} = n\Lambda_{(i)}\wedge\sum  c^{a}C_{a} \, 
\end{equation}
spontaneously breaks $SU(n)$ along the $i$th Cartan direction.  The minimal instanton number is then
\begin{equation} 
\frac{1}{8\pi^2}\int G^{(i)}\wedge G^{(i)} = \frac{n^{2}}{2}(A^{-1})^{ii} = n\sum_{j=1}^{n-1}(A^{-1})^{ij} \, ,
\end{equation}
where in the second equality we have used the group theory identity
\begin{equation}
\sum_{j=1}^{n-1}(A^{-1})^{ij} = \frac{n}{2}(A^{-1})^{ii} \, .
\end{equation}
Therefore the instanton number required to spontaneously break $SU(n)$ in the $i$th Cartan direction is just the sum of the integer coefficients appearing in the $i$th row of the inverse Cartan matrix. For other gauge groups this is not the case, however the number of instantons will always be an integer multiple of the diagonal elements $(A^{-1})^{ii}$.

\subsection{Anomaly freedom is preserved}

Spontaneous symmetry breaking takes {\red one} anomaly-free vacuum into {\red another.} 
Thus we expect that the phase transition of D3-branes into $G$-flux, while giving rise to further gauge symmetry breaking, preserves anomaly freedom. 

We {\red verify this for} the above $SU(5)$ example. 
Although we have as many matter curves as the dimension of the representation, {\red each one} gives the same intersection number. Thus we may {\red choose} the highe{\red st} weight representations and construct the matter surfaces ${\cal S}_{\bf \overline 5} $ and ${\cal S}_{\bf 10}$ as in the following table. \\

\begin{center}
\small
\vspace{-0.5cm}
\begin{tabular}{lll}
\hline \hline
reprs. & highest weight & matter surfaces \\
\hline
 $ {\bf \overline 5} $& $ \mu_{\bf \overline 5} = [0,0,0,1]$ & \parbox{8cm}{ $ {\cal S}_{\bf \overline 5} \equiv E_2 \cdot (3 c_1(B') + \sigma - E_1 -2E_2 -3E_3-2 E_4)$ \\ \text{ \quad \: \, } $ - (8c_1(B')-5B) \cdot (E_1+E_2)$  } \\
 $ {\bf 10} $ &$ \mu_{\bf 10} =[0,1,0,0]$ 
 	& ${\cal S}_{\bf 10} \equiv E_2 \cdot E_4  + ( E_1 + E_2 + E_3 )\cdot c_1(B')$\\
 \hline
\end{tabular}
\end{center}
Accordingly, {\red the chirality is} 
\begin{align}
 \chi({\bf \overline 5})  & = \int_{{\cal S}_{\bf \overline 5}} G_{\,\lambda}^{SU(5)} = 5 \lambda \eta \cdot (\eta -5 B) \, , \\
 \chi({\bf 10}) & = \int_{{\cal S}_{\bf 10}} G_{\,\lambda}^{SU(5)} = 5 \lambda \eta \cdot (\eta -5 B) \, .
\end{align}
Thus we have {\red an} anomaly free theory{\red, since}
\be \label{AnomFreeSU5}
 - \chi({\bf \overline 5}) +  \chi({\bf 10})  = 0 \, .
\ee

{\red Now} assume that a number of D3-branes {\red become} additional flux{\red,} $G^{(4)} = \Lambda^{(4)} \wedge \cal F${\red,} as in (\ref{FourthDir}). The gauge symmetry is broken to $SU(4) \times U(1)$ and the vector multiplet branches {\red as}
\be \label{Bulksu4}
 {\bf 24} \to {\bf 15}_0 + {\bf 4}_{-5} + {\bf \overline 4}_5 + {\bf 1}_0 \, .
\ee
While the diagonal elements remain as vector multiplets, the ${\bf 4}_{-5}$ component becomes chiral {\red according} to (\ref{RRH}), 
\be \label{BulkChirality}
 \chi({\bf 4}_{-5} ) = \int_B c_1 \wedge {\cal F} \, ,
\ee
where $c_1 \equiv c_1(B)$ as before.
{\red The matter fields also branch as}
\begin{align*}
 {\bf \overline 5} & \to {\bf \overline 4}_{1} + {\bf 1}_{-4}, \\
 {\bf 10}& \to {\bf \overline 6}_{-2} + {\bf 4}_{3}. 
\end{align*}
We have the following weight vectors {\red and} corresponding matter surfaces{\red .} 
\begin{center}
\small
\begin{tabular}{lll}
\hline \hline
reprs. & highest weight & matter surfaces \\
\hline
$  {\bf \overline 4}_{1} $& $ \mu_{\bf \overline 5}$ & $  {\cal S}_{\bf \overline 5}$ \\
$   {\bf 1}_{-4}$ & $ \mu_{\bf \overline 5} - \alpha^{(1)}- \alpha^{(2)}- \alpha^{(3)}- \alpha^{(4)}$  
	& ${\cal S}_{\bf \overline 5} + (E_1+E_2+E_3 +E_4) \cdot (8c_1(B') -5 B)$ \\
 $ {\bf \overline 6}_{-2} $ &$ \mu_{\bf 10}$ & ${\cal S}_{\bf 10}$\\
 $  {\bf 4}_{3}$ & $ \mu_{\bf 10} - \alpha^{(2)}- \alpha^{(3)}- \alpha^{(4)}$
 	& ${\cal S}_{\bf 10} - (E_2 + E_3 + E_4) \cdot  c_1 (B')$ \\
 \hline
\end{tabular}
\end{center}
{\red Note} that the first term of {\red each} matter surface is that of the mother representation of the unified group, {\red since} all of them belong to the same highest weight module.  

We obtain the chirality for the representation ${\bf \overline 4}_{1}$ as
\be
 \begin{split}
  \chi({\bf \overline 4}_{1}) & = \int_{{\cal S}_{\bf \overline 5}} G^{(4)}_{\rm tot} = {\cal S}_{\bf \overline 5} \cdot G_{\lambda} +  {\cal S}_{\bf \overline 5} \cdot \Lambda^{(4)} \cdot {\cal F}\\
  & = \chi({\bf \overline 5}) + \frac{1}{5} \int_B (8c_1(B') -5 B) \wedge {\cal F} \\
  &= \chi({\bf \overline 5}) + \frac{1}{5} \int_B (8c_1 - 3t)\wedge {\cal F} \, ,
  \end{split}
\ee
{\red where we have} used the restriction relations $c_1(B')|_B = c_1- t$ and $B|_B = -t$ inside $B'$.

{\red Owing to the fact that} $G^{SU(5)}_\lambda \cdot E_i = 0, i=1,2,3,4$, each {\red matter surface} inherits the chirality of the mother representation.
Note that the matter suface ${\cal S}_{\bf \overline 5}$ is not orthogonal to the divisor $\Lambda^{(4)}$, {\red so} there is {\red a} local chirality change induced by the flux. This contribution is proportional to the $U(1)$ charge, as {\red expected}.
For the representation ${\bf 1}_{-4}$, we have also an additional contribution from the $\alpha^{(4)}${\red, giving}
\be
 \begin{split}
  \chi( {\bf 1}_{-4}) &= \chi({\bf \overline 5}) + \frac{1}{5} \int_B (8c_1 - 3t)\wedge {\cal F} -\int_B (8c_1 - 3t)  \wedge {\cal F} \\
  & = \chi({\bf \overline 5}) - \frac{4}{5} \int_B (8c_1 - 3t) \wedge {\cal F},
  \end{split}
\ee
{\red such that} the additional term {\red is} proportional to the $U(1)$ charge.
Likewise{\red,}
\begin{align*}
 \chi({\bf \overline 6}_{-2})& = \chi({\bf 10}) - \frac{ 2}{5} \int_B c_1(B') \wedge {\cal F} = \chi({\bf 10}) - \frac{ 2}{5} \int_B (c_1-t) \wedge {\cal F} , \\
 \chi( {\bf 4}_{3}) & = \chi({\bf 10}) + \frac{3}{5} \int_B (c_1-t) \wedge {\cal F} .
\end{align*}
Thus{\red, with the contribution from the bulk fermion ${\bf 4}_{-5}$ in (\ref{BulkChirality}), we may verify that there is no anomaly in the $SU(4)$ vacuum.} 
Since the representations $\bf 6$ and $\bf 1$ are real and do not take part in the anomaly, we {\red obtain}
\be \begin{split}
- \chi({\bf \overline 4}_{1}) & + \chi( {\bf 4}_{3}) + \chi( {\bf 4}_{-5}) \\
 =  & -  \chi({\bf \overline 5}) - \frac15 \int_B (8 c_1- 3 t) \wedge {\cal F} + \chi({\bf 10}) + \frac{3}{5} \int_B (c_1-t) \wedge {\cal F}  +  \int_B c_1 \wedge {\cal F}  \\
= &\, 0 \, ,
 \end{split}
\ee
using the relation (\ref{AnomFreeSU5}).

{\red We may also consider $SU(5)$ breaking into $SU(3)\times SU(2) \times U(1)$ under the phase transition induced by $G^{(3)} = \Lambda^{(3)} \wedge {\cal F}$.  The $SU(5)$ representations branch as}
\begin{align*}
 {\bf 24} & \to {\bf (8,1)}_0 + {\bf (1,3)}_0 + {\bf (1,1)}_0 + {\bf (3,2)}_{-5/6} + {\bf (\overline 3,2)}_{5/6}, \\
 {\bf \overline 5} & \to {(\bf \overline 3,1)}_{1/3} + {(\bf1, 2)}_{-1/2}, \\
 {\bf 10}& \to {\bf (3,2)}_{1/6} + {\bf (\overline 3,1)}_{-2/3} + {\bf (1,1)}_{1}. 
\end{align*}
{\red The matter surfaces are as follows.}
\begin{center}
\small
\begin{tabular}{lll}
\hline \hline
reprs. & highest weight & matter surfaces \\
\hline
$   {(\bf \overline 3,1)}_{1/3} $& $ \mu_{\bf \overline 5}$ & $  {\cal S}_{\bf \overline 5}$ \\
$ {(\bf1, 2)}_{-1/2},$ & $ \mu_{\bf \overline 5} - \alpha^{(1)}- \alpha^{(2)}- \alpha^{(3)}$ 
	 & ${\cal S}_{\bf \overline 5} + (E_1+E_2+E_3) \cdot (8c_1(B') -5 B)$ \\
 $ {\bf (\overline 3,1)}_{-2/3} $ &$ \mu_{\bf 10}$ & ${\cal S}_{\bf 10}$\\
 $  {\bf (3,2)}_{1/6}$ & $ \mu_{\bf 10} - \alpha^{(1)}- \alpha^{(2)}- \alpha^{(3)}$ 
 	& ${\cal S}_{\bf 10} - (E_1 + E_2 + E_3)\cdot c_1(B')$ \\
 $ {\bf (1,1)}_{1}$ & $\mu_{\bf 10}- \alpha^{(1)}-2 \alpha^{(2)}- 2\alpha^{(3)}- \alpha^{(4)} $ 
 	& ${\cal S}_{\bf 10} - (E_1+2E_2+ 2E_3+E_4)\cdot c_1(B')$   \\
 \hline
\end{tabular}
\end{center}
Thus the chiralities {\red are}
\begin{align*}
\chi({\bf ( 3,2)}_{-5/6})& =  \int_B c_1 \wedge {\cal F} \, , \\
 \chi( {(\bf \overline 3,1)}_{1/3})  & = \chi({\bf \overline 5}) + \frac{2}{5} \int_B (8c_1-3t) \wedge {\cal F} \, , \\
 \chi( {(\bf1, 2)}_{-1/2}) & =  \chi({\bf \overline 5}) - \frac{3}{5} \int_B (8c_1-3t) \wedge {\cal F} \, , \\
 \chi({\bf (\overline 3,1)}_{-2/3})& = \chi({\bf 10}) - \frac{4}{5} \int_B (c_1-t) \wedge {\cal F} \, , \\
 \chi( {\bf (3,2)}_{1/6}) & = \chi({\bf 10}) + \frac{1}{5} \int_B (c_1-t) \wedge {\cal F} \, , \\
 \chi({\bf (1,1)}_{1}) & = \chi({\bf 10}) + \frac{6}{5} \int_B (c_1-t) \wedge {\cal F} \, .
\end{align*}
The total $SU(3)$ anomaly is {\red therefore}
\be \begin{split} 
 - \chi({(\bf \overline 3,1)}_{1/3}) & - \chi({\bf (\overline 3,1)}_{-2/3}) +2 \chi( {\bf (3,2)}_{1/6}) +2\chi({\bf ( 3,2)}_{-5/6})\\
 &= 
 - \chi({\bf \overline 5}) - \frac{2}{5} \int_B (8c_1-3t) \wedge {\cal F} - \chi({\bf 10})+\frac{4}{5} \int_B (c_1-t) \wedge {\cal F}   \\
 & \quad +2 \chi({\bf 10})+ 2 \cdot \frac{1}{5} \int_B (c_1-t) \wedge {\cal F}+ 2  \int_B c_1 \wedge {\cal F} \\
  & = 0 \, .
\end{split}
\ee
Again, {\red we find a theory with highly non-trivial anomaly cancellation.}

We may {\red therefore} understand how D3-branes {\red contribute indirectly to} the chirality. Their conversion to $G$-flux may affect {\red the chirality locally, but the global sum of chiralities measured by the anomaly remains} zero.

\subsection*{Acknowledgements}
This work is partly supported by the grant NRF-2018R1A2B2007163 of National Research Foundation of Korea.

\end{document}